\documentclass[prl,twocolumn,showpacs]{revtex4}
\usepackage{graphicx}
\begin{document}

\title{Helicoidal ordering in iron perovskites}
\author{Maxim Mostovoy}
\affiliation{Max-Planck-Institut f\"ur Festk\"orperforschung,\\
Heisenbergstrasse~1, D-70569 Stuttgart, Germany}
\date{August 31, 2004}







\begin{abstract}

We consider magnetic ordering in materials with negative charge
transfer energy, such as iron perovskite oxides. We show that for
a large weight of oxygen holes in conduction bands, the double
exchange mechanism favors a helicoidal rather than ferromagnetic
spin ordering both in metals, e.g. SrFeO$_3$ and insulators with a
small gap, e.g. CaFeO$_3$. We discuss the magnetic excitation
spectrum and effects of pressure on magnetic ordering in these
materials.

\end{abstract}

\maketitle

\looseness = -1 The interaction of itinerant electrons with
localized spins couples transport to magnetism and gives rise to
many fascinating phenomena, e.g., the heavy fermion and Kondo
insulator behavior. In manganese perovskites the strong Hund's
rule coupling between the $e_g$ and $t_{2g}$ electrons results in
the colossal magnetoresistance near the metal-insulator transition
\cite{Tokura}. The transport and magnetic properties of manganites
are furthermore strongly influenced by the instability of the
partially filled $e_g$ orbitals towards an orbital ordering.

One would expect to find a similar behavior in other transition
metal (TM) oxides with partially filled $e_g$ and $t_{2g}$ levels.
However, the iron perovskites SrFeO$_3$ and CaFeO$_3$ have very
different properties. Although the tetravalent iron has the same
$t_{2g}^3e_g^1$ electronic configuration as the Mn$^{3+}$ ion in
LaMnO$_3$, the ferrates, which lie closer to the borderline
separating metals and insulators than manganites, exhibit no
Jahn-Teller instability: SrFeO$_3$ is a cubic metal down to lowest
temperatures \cite{MacChesney}, while CaFeO$_3$ is a
charge-ordered insulator below $290$K \cite{Takano}. Both the
metallic and insulating ferrates show the helicoidal magnetic (HM)
ordering with a small helix wave vector ${\bf Q}$ along the body
diagonal, which sets in at $134$K in SrFeO$_3$ and at $115$K in
CaFeO$_3$ \cite{Takeda1,Oda,Woodward}. A further evidence for the
decoupling of transport and magnetism in these materials is the
absence of a resistivity anomaly at N\'eel temperature in
SrFeO$_3$ \cite{Lebon}.

A possible explanation of the suppression of orbital ordering in
Fe$^{4+}$ oxides came from photoemission experiments
\cite{Bocquet}, which showed that SrFeO$_3$ and CaFeO$_3$ have a
large negative charge transfer energy $\Delta \sim -3$eV (i.e.,
the energy necessary to transfer an electron from an oxygen to a
TM ion). For large negative $\Delta$, the conduction bands are
formed by the strongly hybridized iron $e_g$ and oxygen
$p_{\sigma}$ orbitals and the nominal $d^4$ state of the iron ions
has a high weight of the $d^5$ state with holes on oxygen sites.
Since the high-spin $d^5$ configuration is non-degenerate, the
orbital ordering is suppressed. Furthermore, for the half filled
$d$-orbital, the charge fluctuations shift from TM sites to Fe-O
and O-O bonds, reducing the Coulomb energy cost of the charge
ordering.

In this paper we discuss the origin of the helicoidal magnetism in
ferrates. Since magnetism and transport in these materials seem to
be largely decoupled, such an ordering cannot result from the
spin-density-wave (SDW) or any other Fermi surface instability.
The Dzyaloshinskii-Moria interaction \cite{Dzyloshinskii},
responsible for the HM ordering in, e.g., MnSi, is forbidden by
symmetry in the cubic SrFeO$_3$. A non-collinear spin ordering can
be a result of the competition between the ferromagnetic (FM)
double exchange (DE) \cite{Zener,Ahasegawa} and the
antiferromagnetic (AFM) superexchange (SE) between the spins of
the $t_{2g}$ electrons, as was discussed by de Gennes in
Ref.~[\onlinecite{deGennes}]. However, even if we leave aside
problems with the phase separation in DE systems \cite{Dagotto},
the HM state resulting from such a competition, only occurs at low
concentrations of charge carriers, for which the kinetic energy of
the $e_g$ electrons is comparable with the small superexchange
energy of the $t_{2g}$ electrons (the same holds for the spiral
state in the $tJ$-model of doped cuprates \cite{ShraimanSiggia}).
In particular, a non-collinear magnetic ordering in
La$_{1-x}$Sr$_x$MnO$_3$ was reported for $x = 0.06-0.09$
\cite{Geck}. Furthermore, the delicate balance between the DE and
SE, necessary to stabilize the HM state, is incompatible with the
insensitivity of magnetism to transport properties (and vice
versa) in SrFeO$_3$ and CaFeO$_3$.

Here we establish a relation between the helicoidal ordering and
suppression of orbital ordering in ferrates. We show that when the
oxygen orbitals are included in the DE model and the density of
oxygen holes is high, the kinetic energy of itinerant electrons is
minimized for a helicoidal spin ordering. This result holds both
for metals and small gap insulators.

{\it The model:} We consider an extended version of the DE model,
which in addition to itinerant $e_g$ electrons and spins of
localized $t_{2g}$ electrons, includes the oxygen
$\sigma$-orbitals that are strongly hybridized with the TM
$e_g$-orbitals. It is convenient to describe states of the model
in terms of holes that can occupy both oxygen and TM sites. By the
holes on iron sites we mean the $e_g$-holes in the high-spin
electronic $d^5$-configuration (the total spin $S = 5/2$). For
infinite Hund's rule coupling, the spin of the $e_g$ hole on the
site $j$ is antiparallel to the local spin ${\bf S}_j$. Therefore,
these holes can be described by the spinless operators
$d_{j\alpha}$, where the index $\alpha = 1,2$ denotes,
respectively, the $3z^2-r^2$ and $x^2-y^2$ orbitals. The holes on
oxygen sites can have both spin projections: $ p_{j \pm b/2} =
\left(
\begin{array}{c}
p_{j \pm b/2\,\uparrow}\\
p_{j \pm b/2\,\downarrow}
\end{array}
\right), $ where $j \pm b/2$ ($b = x,y,z$) are the $6$ oxygen
sites from the octahedron centered at the TM site $j$.

The Hamiltonian of the DE $dp$-model has the form
\begin{eqnarray}
H_{dp} &=& \sum_{j \alpha b} t_{\alpha b}
\left(d_{j\alpha}^{\dagger} u_j^{\dagger} P_{jb} +
P_{jb}^{\dagger} u_j d_{j\alpha}\right) + t_{pp} \sum_{j,b \neq c}
P_{jb}^{\dagger} P_{jc} \nonumber \\ &+& \Delta \sum_{jb}
p_{j+b/2}^\dagger p_{j+b/2}, \label{eq:Ham1}
\end{eqnarray}
where the $dp-$ and $pp$-hopping amplitudes are expressed through
the Slatter-Koster parameters by $t_{1} = (pd\sigma)
\left(-\frac{1}{2},-\frac{1}{2},1\right)$, $t_{2} = (pd\sigma)
\left(\frac{\sqrt{3}}{2},-\frac{\sqrt{3}}{2},0\right)$, and
$t_{pp} = \frac{1}{2}(pp\sigma) - \frac{1}{2}(pp\pi)$. Since for
infinite Hund's rule coupling the hopping on the TM site $j$ is
only possible for holes with the spin antiparallel to the local
spin ${\bf S}_j = S(\sin\theta_j \sin\phi_j,\cos\theta_j
\sin\phi_j,\cos\theta_j)$, the oxygen hole operator $P_{jb} =
p_{j+b/2} + p_{j-b/2}$ is projected on the spinor
\[
u_{j} = \left(
\begin{array}{c}
-\sin \frac{\theta_j}{2}e^{i\phi_j}\\
\cos \frac{\theta_j}{2}
\end{array}
\right).
\]

For HM state with the wave vector ${\bf Q}$ and spin rotation axis
${\bf {\bf e_3}}$ the local spin on the site $j$
\begin{equation}
{\bf S}_j = S\left({\bf e}_1 \cos {\bf Q} {\bf x}_j + {\bf e}_2
\sin {\bf Q} {\bf x}_j\right), \label{helix1}
\end{equation}
where the unit vectors ${\bf e}_1$, ${\bf e}_2$, and ${\bf e}_3$
form an orthogonal basis. Due to the invariance of the
tight-binding Hamiltonian (\ref{eq:Ham1}) under an arbitrary
rotation of the spin axes, the energy of the HM state is
independent of ${\bf e}_3$. In Sr(Ca)FeO$_3$ the spin rotation
axis is parallel to the helix wave vector \cite{Oda,Woodward} due
to the anisotropy of spin interactions, which we neglect here. For
${\bf e}_3 = {\hat x}$ and ${\bf e_1} = {\hat z}$, we have $ u_{j}
= e^{ \frac{i}{2} \sigma_x \left({\bf Q} {\bf x}_j\right)}
\left(\begin{array}{c} 0
\\ 1 \end{array}\right)$.

To obtain the hole energies for the HM state, the transformation
to the momentum space has to be combined with the spin rotation:
\[
p_{j \pm b/2} = \frac{1}{\sqrt{N}} \sum_{{\bf k}} e^{\left( {\bf
k} + \frac{1}{2} \sigma_x  {\bf Q} \right) {\bf x}_{j \pm b/2}}
p_{{\bf k}b},
\]
where $N$ is the number of Fe ions. The Hamiltonian
(\ref{eq:Ham1}) then reads
\begin{eqnarray}
H &=& \sum_{{\bf k} \alpha b} t_{\alpha b} \left(d_{{\bf
k}\alpha}^{\dagger} P_{{\bf k} b\downarrow} + P_{{\bf
k}b\downarrow}^{\dagger} d_{{\bf k}\alpha}\right) + t_{pp}
\sum_{{\bf k},b \neq c} P_{{\bf k}b}^{\dagger} P_{{\bf k}c}
\nonumber \\&+& \Delta \sum_{{\bf k}b} p_{{\bf k}b}^\dagger
p_{{\bf k}b}, \label{eq:Ham2},
\end{eqnarray}
where
\[
P_{{\bf k}b\sigma} = 2\left( \cos\frac{Q_b}{4} \cos \frac{k_b}{2}
p_{{\bf k}b\sigma} - \sin \frac{Q_b}{4} \sin \frac{k_b}{2} p_{{\bf
k}b,-\sigma} \right) \label{eq:P}
\]
(the lattice constant equals $1$).

\begin{figure}
\centering
\includegraphics[width=7cm]{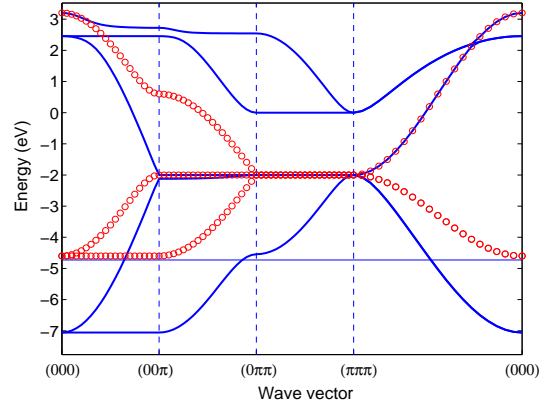}
\caption{\label{fig:bands} The mixed $dp\downarrow$-hole bands
(solid lines) and oxygen $p\uparrow$-hole bands (circles) for
ferromagnetically ordered local spins, $(pd\sigma) = 1.7$eV,
$t_{pp} = 0.65$eV, and $\Delta = -2$eV. The thin horizontal line
indicates the Fermi level for 1 $e_g$-hole/Fe. }
\end{figure}

The hole bands for the FM state are shown in Fig.~\ref{fig:bands}.
For local spins oriented up, the spin-down bands (solid lines) are
formed by the mixed $d-$ and $p-$hole states, while the spin-up
bands (circles) are purely oxygen-hole bands. To a good
approximation the states forming the two lowest bands are Bloch
superpositions of the Zhang-Rice states of holes on metal-oxygen
octahedra \cite{ZhangRice} with the symmetry of the $3z^2-r^2$ and
$x^2-y^2$ orbitals. These are the two bands of the effective
$dd$-model used to describe colossal magnetoresistance manganites
\cite{Dagotto}. The Hilbert space of this model is reduced to the
subspace of the TM orbitals. Even for ferrates with a large
negative $\Delta$, the Fermi sea is either completely or
predominantly filled by the states from these two bands (in
Fig.~\ref{fig:bands} the Fermi energy $\varepsilon_F$ is indicated
by the thin horizontal line).

Yet, for negative $\Delta$ the $dd$-model may fail to describe the
magnetic ground state. The reason is the high density of low
energy spin-flip excitations, created by promoting a hole from the
spin-down Fermi sea to one of the two lowest oxygen spin-up bands,
which at the $\Gamma$-point also have the symmetries of the
$3z^2-r^2$ and $x^2-y^2$ orbitals (see Fig.~\ref{fig:bands}). In
the HM state these bands become mixed with the two lowest
$dp$-bands, which lowers the energies of the occupied states.
There is also an energy loss due to the narrowing of bands caused
by the relative rotation of local spins (the usual DE mechanism).
When the bottom of the oxygen spin-up bands (at $\varepsilon =
\Delta-4t_{pp}$) is close to $\varepsilon_F$, the energy gain
exceeds the energy loss, which makes the FM state unstable and
stabilizes the HM state.

Although both the instability of the FM state towards the HM
ordering and SDW instabilities are driven by the lowering of the
energy of occupied states, the former does not require a nested
Fermi surface and, in general, does not open a gap, since the
crossing of the spin-up and spin-down bands at the Fermi surface
only occurs at isolated points. While the energy driving an SDW
transition is gained close to $\varepsilon_F$, in our case the
energy gain is distributed over the whole Fermi sea - it is a
Fermi sea rather than a Fermi surface instability.

\begin{figure}
\centering
\includegraphics[width=8.6cm]{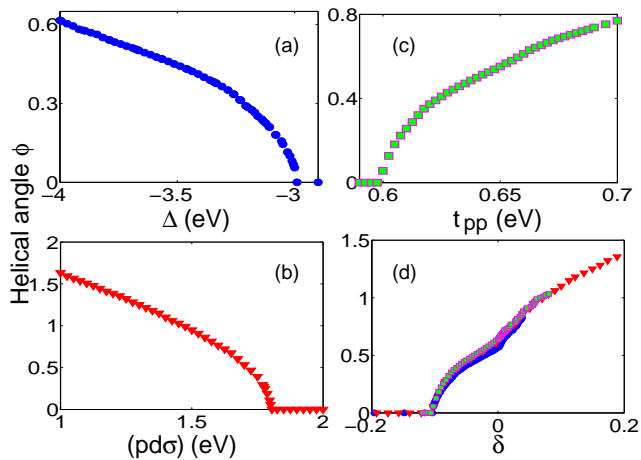}
\caption{\label{fig:threephi} The helical angle $\phi$ (in
radians) between the spins in neighboring $[111]$ layers plotted
versus: (a) $\Delta$, for $(pd\sigma)=1.8$eV and $t_{pp}=0.6eV$;
(b) the hopping amplitude $(pd\sigma)$, for $\Delta = -3$eV and
$t_{pp} = 0.6$eV; (c) $t_{pp}$, for $(pd\sigma)=1.8$eV and $\Delta
= -3$eV. These three curves are re-plotted versus the separation
between the Fermi energy and the bottom of the oxygen band in the
FM state $\delta = (\varepsilon_F - \Delta + 4t_{pp})/(pd\sigma)$
in panel (d).}
\end{figure}

\begin{figure}
\centering
\includegraphics[width=5.5cm]{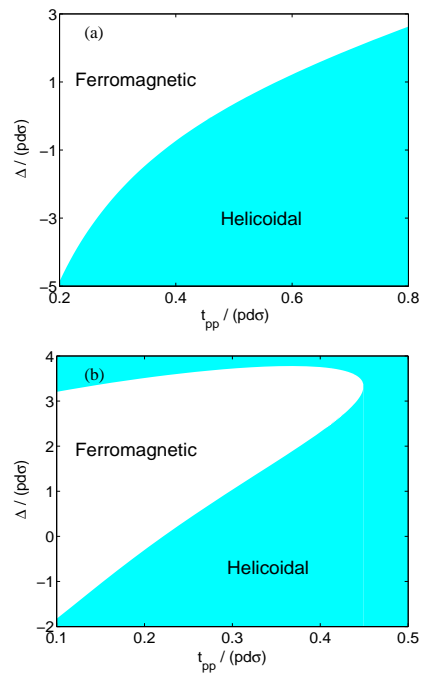}
\caption{\label{fig:phase} The phase diagram of the $dp$-model (a)
and the $dp$-model with the SE interaction (b), for
$\frac{3JS^2}{(pd\sigma)}$ = 0.07.}
\end{figure}

The transition from the FM to HM ground state, induced by varying
$\Delta$, $(pd\sigma)$, and $t_{pp}$, is shown
Fig.~\ref{fig:threephi}, where we plot the optimal angle $\phi$
between the spins in neighboring $[111]$ layers ($\phi = 0$
corresponds to the FM state).  In general, the helical angle
$\phi$ grows with the weight of oxygen holes in the ground state,
which can be achieved by decreasing $\Delta$ (see
Fig.~\ref{fig:threephi}a) or the width of the hybridized
$dp$-bands (see Fig.~\ref{fig:threephi}b), or by increasing the
width of the oxygen bands (see Fig.~\ref{fig:threephi}c). To a
good approximation the value of $\phi$ is controlled by a single
parameter $\delta = \left(\varepsilon_F - \Delta +
t_{pp}\right)/(pd\sigma)$, describing the separation between the
Fermi energy and the bottom of the oxygen spin-up band in the FM
state. When the data shown in Figs.~\ref{fig:threephi} a, b, and c
are re-plotted versus $\delta$, they nearly fall on a single curve
(see Fig.~\ref{fig:threephi}d). The FM state becomes unstable for
small negative $\delta$, when the bottom of the oxygen band is
slightly above the Fermi energy.
The phase diagram of the $dp$-model Eq.(\ref{eq:Ham1}) is plotted
in Fig.~\ref{fig:phase}a. For large $-\frac{\Delta}{(pd\sigma)}$
and $\frac{t_{pp}}{(pd\sigma)}$, the HM state has a lower energy
than the FM one. The ratio of $\Delta$ and $(pd\sigma)$ can be
decreased by pressure, which widens the conduction bands.
Experimentally the switching from HM to FM state was observed in
SrFeO$_3$ at $\sim 13$GPa \cite{Nasu}.

{\it Effect of superexchange:} The AFM superexchange,
\begin{equation}
H_{SE} = J \sum_{ib} \left({\bf S}_i {\bf S}_{i+b}\right),
\;\;\;\;J
> 0, \label{AFM}
\end{equation}
between the spins of the $t_{2g}$ electrons stabilizes a HM state
also at large $\frac{\Delta}{(pd\sigma)}$ (see
Fig.~\ref{fig:phase}b), since for $\Delta \gg (pd\sigma)$,
$H_{dp}+H_{SE}$ reduces to the DE model considered by de Gennes
\cite{deGennes}. We believe, however, that the HM state found in
SrFeO$_3$ lies below the FM region rather than above it. The
`upper' helicoidal state only appears for very large positive
$\Delta$. Furthermore, $\frac{3JS^2}{(pd\sigma)} = 0.07$ used to
obtain Fig.~\ref{fig:phase}b, is too large for SrFeO$_3$. The
applicability of the de Gennes mechanism to SrFeO$_3$ would
require a strong reduction of hopping amplitudes, e.g., due to
polaronic or strong correlation effects.

\begin{figure}
\centering
\includegraphics[width=6cm]{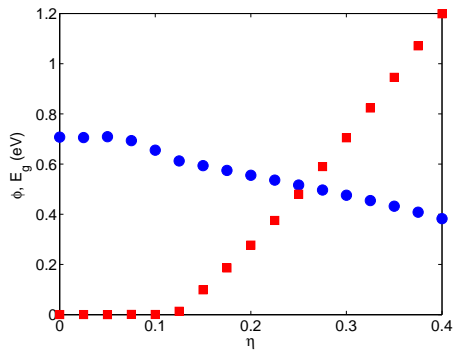}
\caption{\label{fig:etaphig} The helical angle $\phi$ (circles),
measured in radians, and the gap $E_g$ (squares), measured in eV,
versus the modulation amplitude $\eta$ for $(pd\sigma)=1.7$eV,
$t_{pp}=0.65$eV, and $\Delta = -3$eV.}
\end{figure}

{\it Magnetic  ordering in insulators:} Since the instability of
the FM state towards HM ordering, is a Fermi sea rather than a
Fermi surface instability, it can also occur in insulators, such
as CaFeO$_3$ below $T_{CO} = 290$K. Here we assume that the charge
ordering in this material occurs due to a strong electron-lattice
coupling, which for one hole/Fe gives rise to a breathing-type
lattice distortion (an alternation of large and small oxygen
octahedra), observed below $T_{CO}$ \cite{Woodward}.
Figure~\ref{fig:etaphig} shows the dependence of the gap in the
hole spectrum (squares) due to the modulation of the hopping
amplitudes in the ordered state: $t_{\alpha b}(i) = \left(1 +
\sigma_i \eta\right) t_{\alpha b}$ and $t_{pp}(i) = \left(1 +
\sigma_i \eta \right) t_{pp}$, where $\sigma_i = \pm 1$ for $i \in
A/B$ sublattice of the cubic lattice. Since the Fermi surface is
not nested, the gap only opens above some critical value of the
modulation amplitude $\eta$ (a finite electron-phonon coupling
constant is necessary to stabilize the charge ordered state). The
helical angle $\phi$ (circles) decreases very slowly with $\eta$
and even the $1$eV gap does not have a strong effect on the wave
vector of the HM ordering.

\begin{figure}
\centering
\includegraphics[width=7cm]{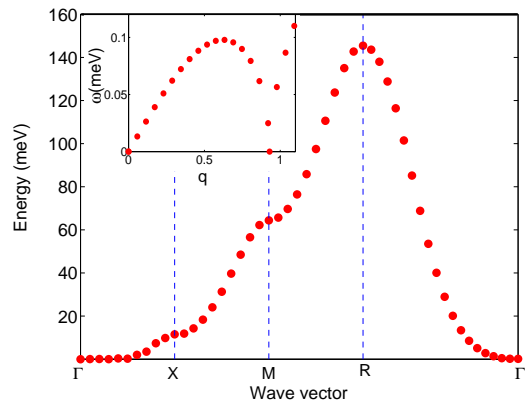}
\caption{\label{fig:hmagnon} The magnon spectrum for the
helicoidal state with ${\bf Q} = 0.93(1,0,0)$, which is the ground
state for $(pd\sigma)=1.7$eV, $t_{pp}=0.65$eV, and $\Delta =
-2$eV. The inset shows $\omega_{\bf q}$ for ${\bf q} = q(1,0,0)$,
where $q$ varies between $0$ and $1.2Q$.}
\end{figure}

{\it Magnon dispersion:} The magnon spectrum for a HM state,
calculated in the leading $1/S$ approximation, is shown in
Fig.~\ref{fig:hmagnon} (details of this calculation will be
published elsewhere). The magnon frequency vanishes both at ${\bf
q} = 0$ and ${\bf q} = {\bf Q}$, corresponding to the two
Goldstone modes: the translation of the incommensurate helix along
${\bf Q}$ and rotation of the helical axis ${\bf e}_3$,
respectively. Although the HM state is stable, the magnon spectrum
for ${\bf q}$ varying between $0$ and ${\bf Q}$ is extremely soft
(note the small energy scale in the inset in
Fig.~\ref{fig:hmagnon}).

At the transition from HM to FM state, these two points merge,
resulting in the vanishing spin stiffness. Thus the transition
from HM to FM state at zero temperature, which can be induced by
varying $\Delta$, $(pd\sigma)$, or $t_{pp}$ (see
Fig.~\ref{fig:threephi}), is a quantum critical point, at which
both the Curie and N\'{e}el temperatures drop to zero due to
diverging spin fluctuations. In reality such a quantum critical
behavior can be suppressed by the single-ion and exchange
anisotropies, neglected in our model, which open a gap in the
magnon spectrum at ${\bf q} = {\bf Q}$ and result in a first order
transition between the two magnetic states. This may explain, why
the magnetic ordering temperature in SrFeO$_3$ monotonously grows
with pressure, even though the HM state is replaced by the FM one
at around $13$ GPa \cite{Nasu}.

In conclusion, we showed that in double exchange systems with a
high density of oxygen holes the kinetic energy of the conduction
electrons is minimized for a helicoidal spin ordering. This allows
us to relate the helicoidal ordering observed in SrFeO$_3$ and
CaFeO$_3$ to the negative charge transfer energy, inferred for
these materials from photoemission experiments.  The difference
between the magnetism in ferrates and manganites is very similar
to the difference between the Mott-Hubbard and charge transfer
insulators \cite{ZaanenSawatzkyAllen}: the lowest energy spin-flip
excitation in ferrates resides mainly on oxygen sites and does not
cost much energy. When the holes predominantly occupy TM sites,
their kinetic energy is minimized for parallel local spins. On the
other hand, the kinetic energy of the oxygen holes is minimal for
the unpolarized Fermi sea. When both the transition metal and
oxygen sites are occupied with high probability, the compromise is
the helicoidal ordering.

I would like to thank B. Keimer, G. Khaliullin, D. Khomskii, O.
Sushkov, B. Simons, and C. Ulrich for fruitful discussions.

\end{document}